\documentclass[12pt,amsmath,amssymb,unsortedaddress,superscriptaddress,prb]{revtex4-1}
\usepackage{graphicx}
\usepackage{dcolumn}
\usepackage{bm}

\begin{document}

\preprint{1}

\title{Drain current modulation in a nanoscale field-effect-transistor channel by single dopant implantation}

\author{B. C. Johnson\footnote{Email: johnsonb@unimelb.edu.au}}
\affiliation{%
School of Physics, University of Melbourne, Victoria 3010, Australia
}%

\author{G. C. Tettamanzi}
\affiliation{%
Kavli Institute of Nanoscience, Delft University of Technology, Lorentzweg 1, 2628 CJ Delft, The Netherlands
}%

\author{A. D. C. Alves}
\author{S. Thompson}
\author{C. Yang}
\affiliation{%
School of Physics, University of Melbourne, Victoria 3010, Australia
}%

\author{J. Verduijn}
\author{J. A. Mol}
\affiliation{%
Kavli Institute of Nanoscience, Delft University of Technology, Lorentzweg 1, 2628 CJ Delft, The Netherlands
}%

\author{R. Wacquez}
\affiliation{%
INAC-SPSMS, CEA-Grenoble, 17 rue des martyrs, F-38054 Grenoble, France
}%

\author{M. Vinet}
\affiliation{%
LETI-Minatec, CEA-Grenoble, 17 rue des martyrs, F-38054 Grenoble, France
}%

\author{M. Sanquer}
\affiliation{%
INAC-SPSMS, CEA-Grenoble, 17 rue des martyrs, F-38054 Grenoble, France
}%

\author{S. Rogge}
\affiliation{%
Kavli Institute of Nanoscience, Delft University of Technology, Lorentzweg 1, 2628 CJ Delft, The Netherlands
}%

\author{D. N. Jamieson}
\affiliation{%
School of Physics, University of Melbourne, Victoria 3010, Australia
}%

\begin{abstract}

We demonstrate single dopant implantation into the channel of a silicon nanoscale metal-oxide-semiconductor field-effect-transistor. This is achieved by monitoring the drain current modulation during ion irradiation. Deterministic doping is crucial for overcoming dopant number variability in present nanoscale devices and for exploiting single atom degrees of freedom. The two main ion stopping processes that induce drain current modulation are examined.  We employ 500~keV He ions, in which electronic stopping is dominant, leading to discrete increases in drain current and 14~keV P dopants for which nuclear stopping is dominant leading to discrete decreases in drain current.

\end{abstract}

\hfill 1

\maketitle

Classical metal-oxide-semiconductor field-effect-transistors (MOSFETs) fabricated by industrial methods are now sufficiently small that random variations in the number and placement of dopants results in inconsistent behaviour. This is already a major issue in the microelectronics industry for devices operating at room temperature.\cite{ITRS:2007wt}  Further, the Bohr radius of a donor electron is now a significant fraction of the device size resulting in the possibility of quantum mechanical dependent functionalities as observed with adventitiously doped devices at 4~K.\cite{LANSBERGEN:2008kx,Pierre:2010uq,klein:043107} Emerging deterministic doping technologies aim to mitigate statistical fluctuations in the doping of these devices while also providing significant potential for solid-state quantum computers.\cite{Kane:1998wj,PhysRevA.62.012306,PhysRevB.67.121301,PhysRevB.74.045311}

Low energy single dopant implantation into micronscale devices has been reported.\cite{Jamieson:2005aa,batra:193502} Further, time-resolved control and transfer of a single electron between two deterministically implanted P atoms has been demonstrated.\cite{Andresen:2007kx} Deterministic doping schemes which employ ion implantation are based on ion impact signals from electron-hole pairs,\cite{Jamieson:2005aa,seamons:043124} secondary electrons\cite{Shinada:2005zt,persaud:2798,Persaud:2004df} or modulation of the drain current, $I_{d}$.\cite{weis:2596,batra:193502,shinada} For the latter, discrete downward steps in $I_{d}$ have been observed with low energy Si implantation into a micronscale SOI wire.\cite{shinada} However, for micronscale MOSFETs, other reports show discrete upward steps.\cite{weis:2596,batra:193502} By an appropriate choice of ion and implant energy we can selectively induce discrete upward or downward steps in $I_{d}$ to elucidate the mechanisms involved in these opposing responses in nanoscale MOSFETs. The full potential of new single-atom functionalities requires nanoscale devices. For example, multi-gate silicon-on-insulator (SOI) transistors are promising architectures.\cite{Tettamanzi:2010kl} 

Here, we examine $I_{d}$ modulation in nanoscale SOI MOSFETs from the passage through the channel of 500~keV He$^{+}$ ions for which electronic stopping is the dominant mechanism for dissipation of the kinetic energy.  We contrast this with the modulation induced by 14~keV P$^{+}$ dopants which mainly stop in the channel and for which nuclear stopping is dominant. In the latter case this modulation is the deterministic signal where precision placement is optimised by using a specialised gate structure which also acts as a surface mask.

\begin{figure}[h!]
\includegraphics[width=15cm]{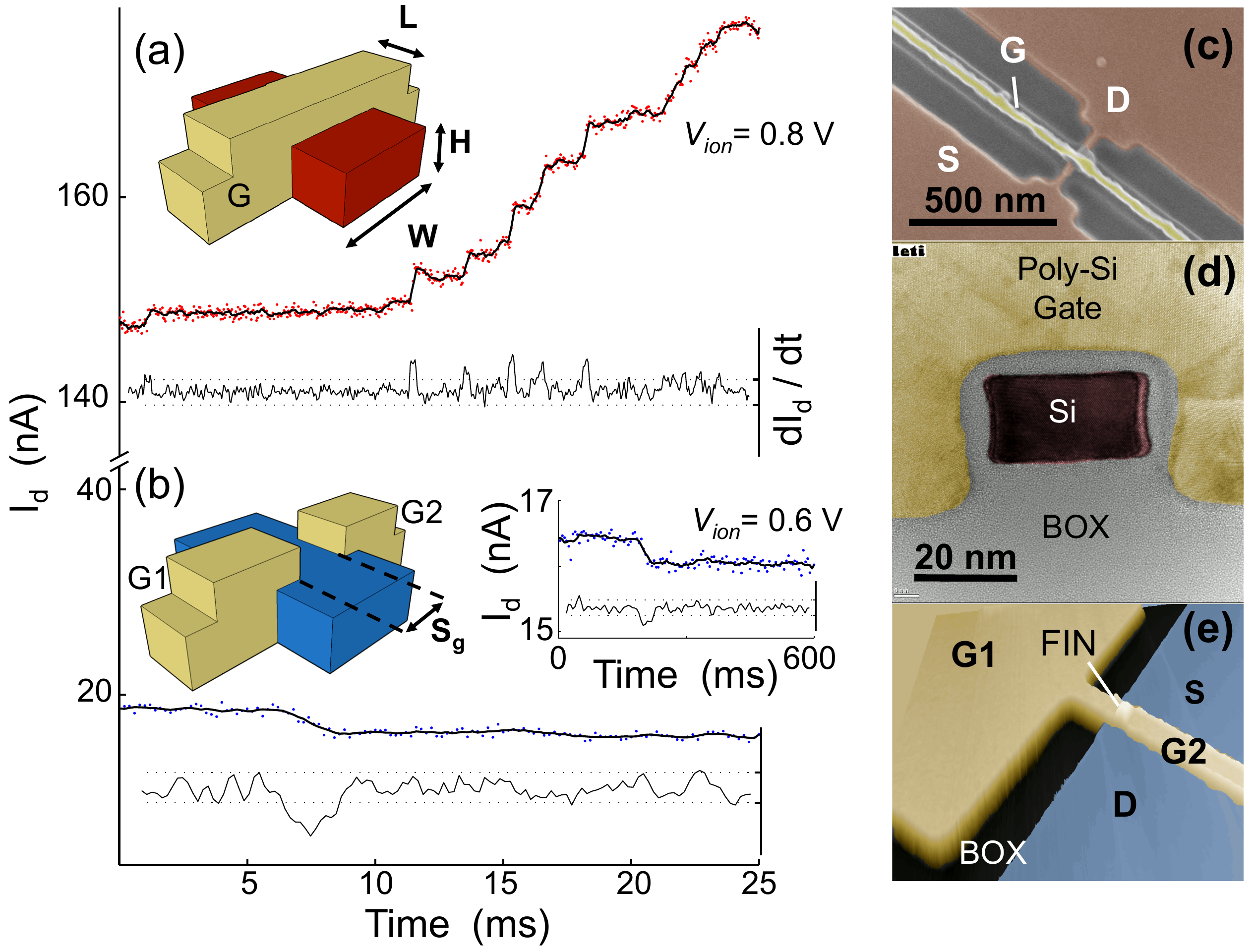}
\caption[]{Nanoscale MOSFET $I_{d}$ collected at a 100~kHz sample rate during (a) 500~keV He and (b) 14~keV P irradiation. Discrete steps represent single ion impacts. The time trace has been binned down to (a) 25~kHz, (b) 5~kHz and 0.2~kHz using the time scale of the step as a guide. The derivative is shown under each trace. Schematics of the devices with channel width ($W$), length ($L$) and height ($H$) are shown. The top right inset of (b) shows the second step observed $\sim$2 minutes after the first step. A false colour SEM image,TEM image of the channel cross section and AFM image of a double gated MOSFET identical to those under study are shown in (c), (d) and (e) respectively.}
\label{itime}
\end{figure}

We fabricated finFETs with SOI consisting of 20~nm of Si on a 145~nm thick buried oxide (BOX). Images of the devices are shown in Figs.~\ref{itime} (c)-(e). The channel had a 5~nm SiO$_{2}$ gate oxide.\cite{pierre:242107} The nominal channel dimensions are listed in Table~\ref{devices}. The SiO$_{2}$/Si interface is expected to have an interface state density in the mid-$10^{10} \;\rm eV^{-1}.cm^{-2}$ as measured by deep level transient spectroscopy. The gates were poly-Si and the source-drain contacts were As doped by ion implantation to a concentration of $\sim$$10^{20} \;\rm cm^{-3}$. Two MOSFET types were considered and are shown schematically in the inset of Fig.~\ref{itime} (a) and (b). The first was a single gate MOSFET with full back-end processing that incorporated a surface passivation layer and was used for the He implantation experiments. The second type was a double gated MOSFET with a spacing between the two gates of $S_{g}=50$~nm. Si$_{3}$N$_{4}$ was formed around each gate leaving a space through which the channel was exposed to the ion beam.\cite{Pierre:2010uq}

The MP2 beam-line\cite{Jamieson1997706} and a Colutron implanter\cite{Jamieson:2005aa} were used to irradiate the devices with 500~keV He$^{+}$ and 14~keV P$^{+}$, respectively. The source, drain and gate were bonded into a chip carrier with electrostatic discharge protection. The He$^{+}$ beam was focused to produce a sharp horizontal line ($3\times 1000\;\rm\mu m^{2}$) that was scanned across the device and had a beam flux of $\sim$$7\times 10^{12} \;\rm ions/cm^2/s$. The P$^{+}$ beam was directed through a stationary 600~$\mu m$ diameter aperture at an average flux of $\sim$$2\times 10^{9} \;\rm ions/cm^2/s$. During implantation, $I_{d}$ was monitored at a gate voltage of $V_{ion}$ which is shown for each device in Table~\ref{devices}. $V_{ion}$ was chosen so that any shift in threshold voltage could be detected with the associated change in $I_{d}$. The IV characteristics of all devices were measured before and a day after the irradiation using standard electronics (Keithley 487).

Figures~\ref{itime} (a) and (b) show the variation of $I_{d}$ during He$^{+}$ and P$^{+}$ irradiations, respectively. Discrete steps in $I_{d}$ are observed in both cases and represented by peaks in the $dI_{d}/dt$ plot. For the He irradiated device there is some variation in the peak height. This can be understood from our PADRE\cite{941} device simulations (not shown) which suggest that the threshold voltage shift is most sensitive where the current density is high. An ion strike where the current density is low will have a smaller effect and an ion strike in the source or drain will have a negligible effect. Generally, the $I_{d}$ step heights become smaller as irradiation continues signifying that the IV curve is shifting to a point where $I_{d}$ does not vary significantly around $V_{ion}$. For the irradiation with P$^{+}$ two discrete steps in $I_{d}$ are observed (Fig.~\ref{itime} (b)). The second step is shown in the top right inset on a different scale. The time constant of this second step is much larger than the first and its height is also much smaller, most likely as a consequence of the transformation of a significant volume of the channel as a result of the previous ion strike. After irradiation, the devices remained robust and there was no observable change in gate leakage current. 

Table~\ref{devices} shows the estimated fluence calculated from the measured beam flux and the implant duration. The total exposure is the number of ions that would be implanted given this fluence and the device dimensions. The counted He$^{+}$ impact signals are reasonably consistent with this estimate. For P$^{+}$, the estimate does not take into account the reduced size of the exposed channel area caused by the Si$_{3}$N$_{4}$, hence the larger discrepancy. $I_{d}$ was found to step up with He$^{+}$ implantation and down with P$^{+}$. After the beam was no longer incident on $He2$, $I_{d}$ began to decrease over a longer time scale. This is indicative of the recombination of positive trapped oxide charge\cite{trombetta:2512} and was not observed for P$^{+}$ implanted devices.

\begin{figure}[h!]
\includegraphics[width=15cm]{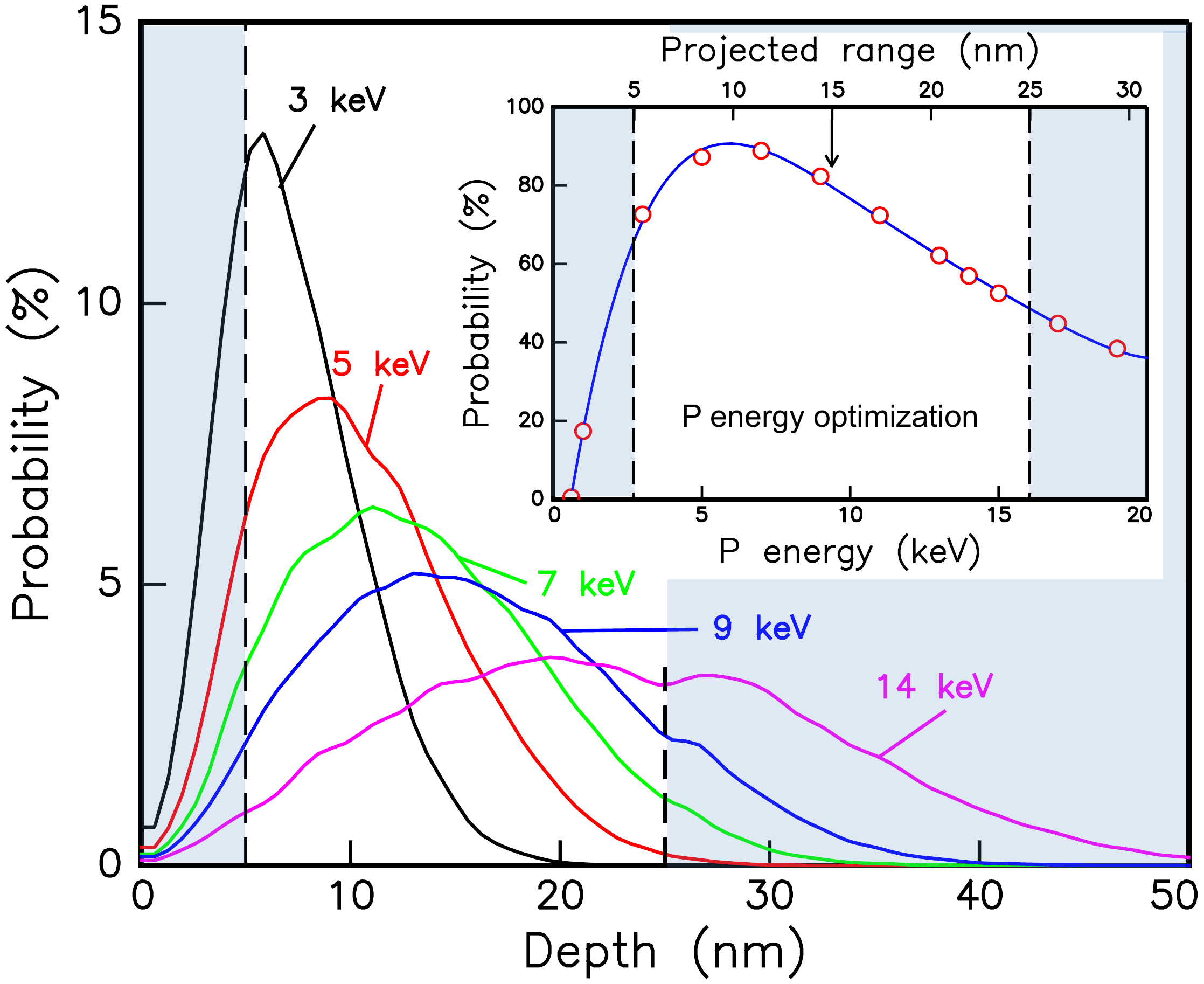}
\caption[]{Probability of single P$^{+}$ donor placement calculated with SRIM\cite{srim} as a function of depth through the nanoscale SOI MOSFET. The inset shows the probability of placing a P$^{+}$ atom in the channel. The shaded area represents the SiO$_{2}$ part of the device. }
\label{srim}
\end{figure}

Monte Carlo SRIM\cite{srim} simulations of P$^{+}$ implants into a simplified nanoscale SOI structure are shown in Fig.~\ref{srim}. While we demonstrate that a discrete number of ions can be implanted, they are subject to random statistical processes that cause straggle in the ion range. For these devices, a 5.9~keV P$^{+}$ is optimimum for donor placement in the channel with a probability of 90\%. We have used 14~keV for direct comparison with earlier work.\cite{Tan:2009uq} This results in a 57\% chance of the P$^{+}$ stopping within the channel. SRIM simulations further show that the concentration of vacancies created by 14~keV P$^{+}$ is an order of magnitude greater than 500~keV He in the channel of the MOSFET. Conversely, the He$^{+}$ causes about twice as many ionisations than the P$^{+}$. This illustrates that the mechanisms by which $I_{d}$ is modified depends greatly on the type and energy of implanted ion. Ionisation can result in trapped positive charge in the BOX.\cite{dekeersmaecker:532} In this work, it is likely that the dominant $I_{d}$ modulation is either caused by ionisations in the BOX for He$^{+}$ and Frenkel pairs created in the channel region for P$^{+}$. 

\begin{figure}[h!]
\includegraphics[width=12cm]{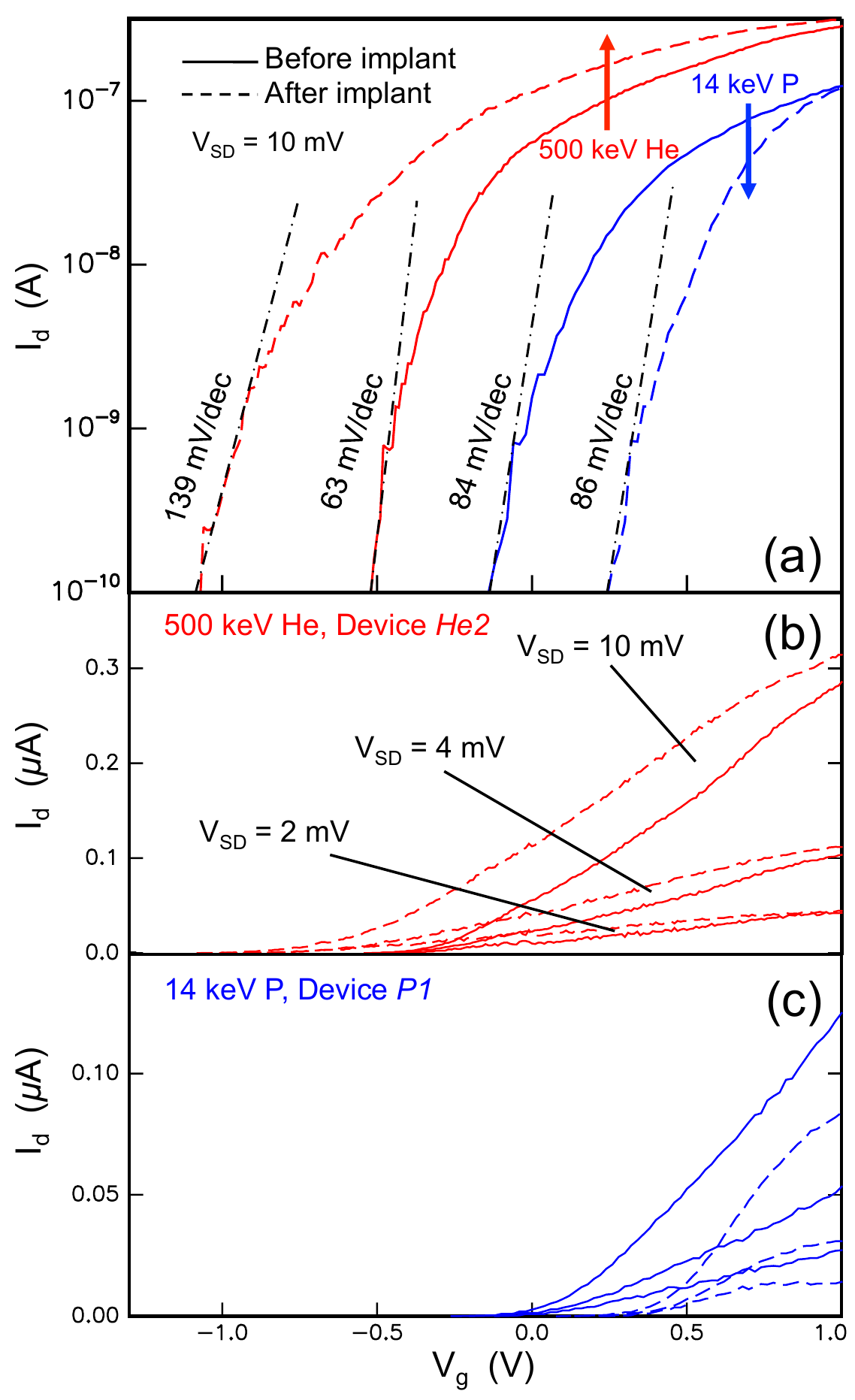}
\caption[]{IV curves before and after irradiation with 14~keV P$^{+}$ (blue) and 500~keV He$^{+}$ (red) showing (a) the subthreshold region and the saturation region for (b) $He2$ and (c) $P1$. The arrows in (a) show the direction $I_{d}$ shifts during implantation.}
\label{sub}
\end{figure}

Figure~\ref{sub} shows the IV curves for devices $He2$ and $P1$ before and one day after the implants. The He$^{+}$ and P$^{+}$ implantation induced defects modify the IV curves in different ways. Similar results are found for $He1$. The trapped oxide charge created in the BOX by the He implants can result in an inversion layer formed along the Si/BOX interface which causes interface coupling effects.\cite{Cristoleveanu:2004uq,Eminente2007239,Daug2004535} This results in the observed negative shift. The charge density was estimated using analytical expressions of the subthreshold $I_{d}$ at midgap to be $5.0 \times 10^{12}\;\rm cm^{-2}$ for both $He1$ and $He2$.\cite{mcwhorter:133} In addition to a shift, there is a noticeable stretch-out, the extent of which is indicated by the subthreshold swing, $S$ in Fig.~\ref{sub} (a). We find the associated effective change in interface trap density is $4.3 \times 10^{12}\;\rm cm^{-2}eV^{-1}$ and $5.5 \times 10^{12}\;\rm cm^{-2}eV^{-1}$ for devices $He1$ and $He2$, respectively. 

The P$^{+}$ implantation caused quite different behaviour as seen in Fig.~\ref{sub}(c). A positive shift is observed suggesting that the interface states are negatively charged as is the case for n-type MOSFETs.\cite{90114} This shift corresponds to a charge density of $1.5 \times 10^{12}\;\rm cm^{-2}$. A decrease in $I_{d}$ is also observed (Fig.~\ref{sub}(b)) suggesting an increase in series resistance consistent with the introduction of Frenkel pairs in the channel.

In conclusion, the implantation of single P$^{+}$ ions into a nanoscale SOI MOSFET was demonstrated. $I_{d}$ was found to depend on the stopping mechanisms and where in the substrate the ion energy was deposited. Electronic stopping resulted in trapped charge in the BOX causing a threshold voltage shift. Conversely, the detection of low energy implanted dopants required for deterministic doping relies on a series resistance increase caused by the introduction of Frenkel pairs into the channel. 

The authors acknowledge Þnancial support from the EC FP7 FET-proactive NanoICT projects MOLOC (215750) and AFSiD (214989) and the Dutch Fundamenteel Onderzoek der Materie FOM.

\newpage

\begin{center}
\begin{table*}[*ht]
\caption[]{\label{devices} Summary of the devices under study.}
\begin{tabular}{c c c c c c c}
\hline \hline
ID & Gate &  $L\times W\times H$ & Estimated 	&Total & Counted & $V_{ion}$\\ 
     & type &   (nm$^{3}$) & fluence (cm$^{-2}$)	& exposure$^{a}$& ions &  (V)\\ 
\hline
$P1$ 	& Double	& $25\times 70\times 20$$^{b}$	& $5\times 10^{12}	$  	& $6\times 10^{1}$	P	& 2 &0.6	\\ 
$He1$	& Single	& $25\times 60\times 20$			& $3\times 10^{12}$		& $5\times 10^{1}$ He		& -		&grounded \\ 
$He2$	& Single	& $45\times 60\times 20$			& $3\times 10^{12}$ 		& $8\times 10^{1}$ He		&30  		& 0.8			\\ 
\hline \hline
\end{tabular}
\end{table*}
\end{center}
{\small $a$ Upper limit of implanted ions into the device subject to Poisson statistics and experimental uncertainties. \\
$b$ The exposed space between the two gates on top of the channel, $S_{g}$, was 50~nm wide. Si$_{3}$N$_{4}$ spacers decrease the exposed area further.}

\newpage

\end{document}